\documentclass[leqno,10pt]{article}
 \usepackage{tikz,graphicx,subfigure,overpic,verbatim}
\usetikzlibrary{arrows,decorations.markings}

\usepackage{amsmath,amsthm,amscd,amssymb}

\usepackage{latexsym}

\usepackage{mathrsfs}

\newcommand{\bbN}{{\mathbb{N}}}

\newcommand{\bbR}{{\mathbb{R}}}

\newcommand{\bbC}{{\mathbb{C}}}

\DeclareMathOperator{\Ai}{{\mathrm Ai}}

\newcommand{\beq}{\begin{equation}}

\newcommand{\eeq}{\end{equation}}

\newcommand{\ba}{\begin{align}}

\newcommand{\ea}{\end{align}}

\DeclareMathOperator{\Real}{Re}
\DeclareMathOperator{\Tr}{Tr}

\DeclareMathOperator{\Ima}{Im}
\renewcommand{\Im}{\Ima}
\renewcommand{\Re}{\Real}

\newcommand{\EE}{{\mathbb{E}}}

\DeclareMathOperator*{\diag}{diag}

\DeclareMathOperator*{\arrow}{\rightarrow}

\allowdisplaybreaks

\numberwithin{equation}{section}

\newtheorem{theorem}{Theorem}[section]

\newtheorem{proposition}[theorem]{Proposition}

\theoremstyle{definition}

\newtheorem{rhp}[theorem]{RH problem}

\theoremstyle{remark}

\title{Painlev\'e kernels in Hermitian matrix models}
\author{Maurice Duits\footnote{Department of Mathematics, Royal Institute of Technology (KTH), Lindstedtsv\"agen 25, SE-10044 Stockholm, Sweden. Supported by the grant KAW 2010.0063 from the Knut and Alice Wallenberg Foundation.}}

\date{}

\begin{document}

\maketitle
\begin{abstract}
  After reviewing the Hermitian one matrix model, we will give a brief introduction to the Hermitian two matrix model and present  a summary of some recent  results on the asymptotic behavior of the two matrix model with a quartic potential. In particular, we will discuss a limiting kernel in the quartic/quadratic case that is constructed out of a $4\times 4$ Riemann-Hilbert problem related to  Painlev\'e II equation. Also an open problem will be presented. \end{abstract}

\section{Introduction}
The study of  the local statistics in large random systems of interacting particles, such as the eigenvalues of large random  matrices, is a central theme in random matrix theory.  The  universality principle  states that   the local statistics  obey laws that have a universal character and do not depend on the precise definition of the underlying probability measure but rather on some general characteristics. 
Generic examples are the sine universality in the bulk and Airy universality near a soft edge. 
However,  there may be singular points near which the limiting local correlations are described by more complicated universality classes. In some of these  cases remarkable connections to  Painlev\'e equations have been  found. 

A  model  on which substantial progress has been made  is the Hermitian one matrix model (also called the Unitary Ensembles). A milestone was the rigorous verification of the bulk and soft edge universality conjectures in \cite{DKMVZ1,DKMVZ2} based on the orthogonal polynomial approach and  Riemann-Hilbert techniques. Among further important developments are the treatment of the three types of singular points that  can occur.  Near two of those  the local correlations are described by $\Psi$-functions associated to special solutions  to members of the Painlev\'e I and II hierarchies \cite{BI,CIK,CK,CV,SH}.

In this paper, our main focus will be on the Hermitian two matrix model which is  a natural extension of the Hermitian one matrix model. There is strong evidence that the two matrix model gives rise to a  family of singular situations that is much larger than the one for the one matrix model. For example, it is believed to be a good model for generating the $(p,q)$ conformal minimal models \cite{DaKaKo}. The  classification and characterization of the local correlations near the singular points is an important open problem in random matrix theory.

By using the Riemann-Hilbert approach for the associated biorthogonal polynomials, we recently analyzed the asymptotic behavior of the two matrix model with one quartic potential \cite{DG,DGK,DK2,DKM1,Mo}. Here we will  report on that  progress.   As it is important to compare our results with the results for the one matrix model, we provide in Section 2 a brief overview of the one matrix model and the singular cases that can appear. In particular, we review the results on the case of one to two interval transition related to the Painlev\'e II equation.  This situation will also return in our discussion on the two matrix model.   

In Section 3 we discuss the definition of the Hermitian two matrix model in its general form and the relation with certain biorthogonal polynomials.   In Section 4 we first  discuss the vector equilibrium problem  that was a key ingredient for the asymptotic analysis \cite{DK2,DKM1,Mo} of the Riemann-Hilbert problem for the biorthogonal polynomials. Second, we present a phase diagram and a new critical phenomenon for the quartic/quadratic case that we analyzed in \cite{DG}.  Interestingly, the limiting kernel is constructed out of a $4\times 4$ Riemann-Hilbert problem associated to the Hastings-McLeod solution to the Painlev\'e II equation.  

Finally, in Section 5 we show that the singular  case  of \cite{DG} can be embedded into a larger class of singular cases  for which a rigorous analysis may be within reach.  We leave this as an open problem.

\section{One matrix model}

The Hermitian one matrix model  is defined as the probability measure  on the space of $n\times n$ Hermitian matrices given by 
	\beq \label{eq:onematrix}
		\frac{1}{Z_n} \exp \left(-n\Tr V(M) \right){\rm d}M,	
	\eeq 
where $V$ is a polynomial of even degree and positive leading coefficient, 
${\rm d}M= \prod_{i=1}^n {\rm d}M_{ii} \prod _{i<j} {\rm d}\Re M_{ij}\, {\rm 	d}\Im M_{ij},$
and $Z_n$ is a normalizing constant.  We will be concerned with the limiting behavior of the eigenvalues of a matrix $M$ taken randomly from \eqref{eq:onematrix}  as $n\to \infty$. 

\subsection{Global limit}
By integrating over the unitary group the probability measure \eqref{eq:onematrix} reduces to the following probability measure on the eigenvalues 
\beq \label{eq:onematrixev}
\frac{1}{\tilde Z_n} \prod_{i<j} (x_i-x_j)^2Ê{\rm e}^{-n \sum_{j=1}^n V(x_j)} \ {\rm d}x_1 \cdots {\rm d}x_n.\eeq
Because of the Vandermonde determinant  the probability of for eigenvalues to cluster is small  and hence the eigenvalues appear to repel each other. On the other hand, due to the exponential factor the probability of finding the eigenvalues spread out is small and hence $V$ acts a  confining potential.  As $n\to \infty$ the eigenvalues seek the equilibrium situation for these two competing mechanisms.  More precisely,  the empirical eigenvalue distribution $\frac{1}{n} \sum_{j=1}^n \delta_{x_j}$ converges weakly to $\mu_V$ almost surely, where $\mu_V$ is the unique minimizer of the energy functional 
 \beq \label{eq:energyexternalfield}
  	I_V(\nu)=\iint \log\frac{1}{|x-y|}{\rm d}\mu(x){\rm d}\mu(y) +\int V(x)Ê{\rm d}\mu(x),
 \eeq 
minimized over all probability  measure $\mu$ on $\bbR$. 

This minimization problem can be solved in terms of  an algebraic curve. There exists a polynomial $Q$ of degree $\deg V-2$ such that the function
$$
\xi(x)= V'(x) -\int \frac{1}{x-s}{\rm d}\mu_V(s),
$$ 
is a solution to the equation
\beq \label{eq:masterloopequation1m}
\xi^2-V'(x)\xi+Q(x)=0.
\eeq 
 This curve can be derived using the variational conditions corresponding to minimizing \eqref{eq:energyexternalfield}. It can also be derived by a change of variables in \eqref{eq:onematrix} (see for example \cite{Jduke}). It particularly implies that  $\mu_V$ is absolutely continuous with a density
$$
\rho(x)=\frac{{\rm d} \mu_V}{{\rm d}x} (x)=\frac1\pi \sqrt{R_-(x)}, \qquad x\in \bbR.
$$
where $R_-$ is the negative part of $R(x)=V'(x)^2/4-Q(x)$ (so $R=R_+-R_-$). In particular, one readily verifies that the support $S(\mu_V)$ of $\mu_V$ consists of a finite  number of intervals. In the generic situation, the density is strictly positive in the interior and vanishes as a square root at the endpoints. 

\subsection{Airy and sine universality}

Although the eigenvalue distribution on the global scale has a deterministic limit, interesting point processes are observed by scaling the eigenvalues around a point in the support of $\mu_V$ such that the average distance between eigenvalues is of finite order. Before we describe the asymptotic behavior on the local scale, we discuss the orthogonal polynomials that integrate the one matrix model. 

For $k\in \bbN_0$ let $p_{k,n}$ be the unique monic orthogonal  polynomial of degree $k$ with respect to ${\rm e}^{-n V(x)}{\rm d}x$ on $\bbR$, i.e.
$$
\int p_{k,n}(x)x^j{\rm e}^{-nV(x)} {\rm d}x =0, \quad j=0,\ldots,k-1.
$$
Moreover, let $K_n$ be the reproducing kernel 
$$
K_n(x,y)={\rm e}^{-n (V(x)+V(y))/2} \sum_{k=0}^{n-1} \frac{1}{h_{k,n}^2}p_{k,n}(x) p_{k,n}(y),
$$
where $h_{k,n}^2=\|p_{k,n}\|_{\mathbb L_2({\rm e}^{-n V(x)} {\rm d}x)}^2$. Then the eigenvalues of $M$ taken randomly with respect to \eqref{eq:onematrix} form a determinantal point process with kernel $K_n$. This means that \eqref{eq:onematrixev} and the marginal densities (or, up to a scalar, the correlation functions) can  be written as 
\beq \label{eq:1mdet}
\underbrace{\int \cdots \int}_{n-k \text{ times }} \mathcal P(x_1,\ldots,x_n){\rm d}x_{k+1}\cdots {\rm d}x_n = \frac{(n-k)!}{n!} \det\left( K_n(x_i,x_j)\right)_{i,j=1}^k,
\eeq
for $k=1,\ldots,n$. 
For more details on determinantal point processes we refer to  \cite{BorDet,HKPV,Jdet,K,L,Sosh}.  The main point is that $K_n$  characterizes the point process entirely and in order to find the asymptotic behavior of the process on the local scale it suffices to compute the scaling limits of the kernel $K_n$.

Let $x^*\in S(\mu_V)$  such that $\rho(x^*)>0$. Then for large $n$ the distance between eigenvalues near $x^*$ is of order $\sim n^{-1}$ and we have the limit
\beq\label{eq:sine}
\lim_{n\to \infty} \frac{1}{n\rho(x^*)} K_n\left(x^*+\frac{x}{n\rho(x^*)},x^*+\frac{y}{n\rho(x^*)}\right)=\frac{\sin \pi(x-y)}{\pi(x-y)}.
\eeq
uniformly for $x,y$ in compact subsets. Note that the right-hand side is independent of $x^*$ and $V$. This is the celebrated principle of sine universality in  random matrix theory.  In \cite{DKMVZ1,DKMVZ2} it was proved by using RH methods for analytic potentials $V$ and  later extended in \cite{McLM,PS} to more general situations. It should also be noted that in recent years there has been an interesting development on proving \eqref{eq:sine} by using more classical methods for orthogonal polynomials, obtaining even more general results. See  \cite{Lubinsky} for a survey.  

Now let  $x^*$ be an endpoint of an interval  in the support of $\mu_V$ and  assume that we are in the generic situation so that the density vanishes as a square root at $x^*$. Then the distance between eigenvalues near $x^*$ is of order $\sim n^{-2/3}$ and we have the scaling limit  
\beq\label{eq:Airy}
\lim_{n\to \infty} \frac{\pm 1}{(c n)^{2/3}} K_n\left(x^*\pm \frac{x}{(c n)^{2/3}},x^*\pm\frac{y}{(c n)^{2/3}}\right)=\frac{\Ai(x)\Ai'(y)-\Ai'(x)\Ai(y)}{x-y}
\eeq
with $+$ is if $x^*$ is a right endpoint and $-$ if $x^*$ is a left endpoint. Here $\Ai$ stands for the Airy function.

\subsection{Singular points}\label{sec:sing}

The scaling limits \eqref{eq:sine} and \eqref{eq:Airy} are valid for the regular points of the spectral curve \eqref{eq:masterloopequation1m}. It may happen that there are singular points and around these points we obtain different and more complicated limits. There are three types of singular points that can occur.
\begin{enumerate}
\item \textbf{Exterior singular point} a point outside the support where we have equality in the variational inequality for the equilibrium problem \eqref{eq:energyexternalfield};
\item \textbf {Interior singular point} a point $x^*$ in the interior of $ S_{\mu_V}$ such that $\rho(x)=c (x-x^*)^{2m}(1+o(1))$ as $x\to x^*$ for some $m\in \bbN$ and $c>0$;
\item \textbf{Singular edge point} an endpoint $x^*$ of one of the intervals in $S_{\mu_V}$ such that $\rho(x)=c (x\pm x^*)^{2m+1/2}(1+o(1))$ as $x\to x^*$ for some $m\in \bbN$  and $c>0$ (here $\pm$ depends whether we have a right or left endpoint).
\end{enumerate} 
 By varying the coefficients of $V$ these singular points correspond to transitions in the support $ S(\mu_V)$. For example, we can let an interval in the support shrink to a point after which  it disappears, leading to an exterior singular  point. By letting two intervals merge to one we create an interior singular point. If both transitions happen simultaneously near the same point we obtain a singular edge point.   By involving several intervals simultaneously we obtain the higher order singularities.
 
  Near singular points,  the local correlations have limits that are different from \eqref{eq:sine} and \eqref{eq:Airy}. The singular edge points and interior singular point are related to special solutions to members of  the Painlev\'e I and Painlev\'e II  hierarchy  respectively, which we will discuss in more detail. The treatment of exterior singular points of the first type can be found in \cite{BL,C,Mo2} but they are  not related to the Painlev\'e transcendents.

\subsubsection{Interior singular points}
We will now discuss the singular case of quadratic vanishing at an interior point in more detail. This situation will also play a role in the upcoming discussion on the  two matrix model. 

Let us consider the double well potential 
$$V(x)=\tfrac14 x^4 - x^2.$$
In this case, the origin is a singular interior point and the density vanishes quadratically
$$\rho(x)=\tfrac{2}{\pi} x^2 \sqrt {2-x^2} {\rm d}x$$
This situation was analyzed in \cite{BI}. For the treatment of more general potentials leading to  quadratic vanishing at an interior point see  \cite{CK,SH}.  

The singular case is analyzed by means of a double scaling limit. We  introduce a parameter $\lambda$ and consider the random matrix model with $V_\lambda=\lambda V$.   For $\lambda<1$ the origin is a regular point and the $S_{\mu_V}$ consists of one interval. For $\lambda>1$ a gap opens up at the origin and the support consists two intervals.   For  the critical value $\lambda=1$ the density of the equilibrium measure vanishes quadratically at the origin. We take the limit $n\to \infty$ but at the same time we let the gap open or close. That is,  we let $n\to \infty$ and $\lambda_n \to 1$ simultaneously such that 
$$n^{2/3}(\lambda_n-1) \to s\in \bbR. $$ Then for some constants $c_1,c_2 >0$ we have  
\beq\label{eq:PII}
\lim_{n\to \infty,\lambda_n\to 1} \frac{1}{(c_1 n)^{1/3}} K_n\left(x^*+ \frac{x}{(c_1 n)^{1/3}},x^* + \frac{y}{(c_1 n)^{1/3}};\lambda_n\right)=K_{\rm PII}(x,y;c_2s).
\eeq
Here $K_{\rm PII}(\cdot,\cdot;\nu)$ with $\nu \in \bbR$ is a family of kernels for which the description is more complicated than the limits in \eqref{eq:sine} and \eqref{eq:Airy}. It can  be characterized in terms of a Lax pair for the Hastings-McLeod solution to the Painlev\'e II equation. Here we will treat the Riemann-Hilbert problem(= RH problem) characterization of the kernel. See \cite{FIKN} for more details on the Riemann-Hilbert approach to the Painlev\'e equations.

First we define the contour $\Sigma_\Psi=\Gamma_1 \cup \Gamma_2 \cup \Gamma_3 \cup \Gamma_4$, consisting of the rays
\[
\Gamma_1=e^{\pi i/6}\bbR^+, \quad \Gamma_2=e^{5\pi i/6}\bbR^+, \quad \Gamma_3=-\Gamma_1, \quad \Gamma_4=-\Gamma_2.
\] 
All rays are oriented towards infinity. The orientation also allows us to provide each point of the contour with a $+$ side lying at the left and a $-$ side lying at the right when traversing the contour according to its orientation. 
\begin{figure}[t]
\centering
\begin{tikzpicture}[scale=1.2]
\begin{scope}[decoration={markings,mark= at position 0.5 with {\arrow{stealth}}}]
\draw[postaction={decorate}]      (0,0)--node[midway, above]{$\Gamma_1$}(1.73,1) node[right]{$J_1=\begin{pmatrix} 1&0 \\ 1&1 \end{pmatrix}$};
\draw[postaction={decorate}]      (0,0)--node[midway, above]{$\Gamma_2$}(-1.73,1) node[left]{$J_2=\begin{pmatrix} 1&0 \\ -1&1 \end{pmatrix}$};
\draw[postaction={decorate}]      (0,0)--node[midway, below]{$\Gamma_3$}(-1.73,-1) node[left]{$J_3=\begin{pmatrix} 1&1 \\ 0&1 \end{pmatrix}$};
\draw[postaction={decorate}]      (0,0)--node[midway, below]{$\Gamma_4$}(1.73,-1) node[right]{$J_4=\begin{pmatrix} 1&-1 \\ 0&1 \end{pmatrix}$};
\draw (1.4,1) node {\tiny{+}};
\draw (1.6,.8) node {{-}};
\draw (-1.4,1) node {{-}};
\draw (-1.6,.8) node {\tiny{+}};
\draw (1.4,-1) node {\tiny{+}};
\draw (1.6,-.8) node {{-}};
\draw (-1.4,-1) node {{-}};
\draw (-1.6,-.8) node {\tiny{+}};\end{scope}
\end{tikzpicture}
\caption{The jump contour $\Sigma_\Psi$ in the complex $\zeta$-plane and the constant jump matrices $J_k$ on each of the rays $\Gamma_k$, $k=1, \ldots,4$.}
\label{fig: contour Psi}
\end{figure}

\begin{rhp} \label{rhp: PII rhp}
For $\nu \in \bbR$, we look for a $2 \times 2$ matrix-valued function $\Psi(\zeta;\nu)$ satisfying
\begin{itemize}
\item[\rm (1)] $\Psi(\zeta;\nu)$ is analytic for  $\zeta \in \bbC \setminus \Sigma_\Psi$;
\item[\rm (2)] $\Psi_+(\zeta;\nu)=\Psi_-(\zeta;\nu)J_k$, for $\zeta \in \Gamma_k,$ $k=1,\ldots,4$. Here $\Psi_\pm$ stands for the limiting value of $\Psi$ at the $\pm$ side of $\Gamma_k$ and $J_k$ is as in Figure \ref{fig: contour Psi}.
\item[\rm (3)] As $\zeta \to \infty$ we have
\[
\Psi(\zeta;\nu)=\left( I+ \mathcal O(\zeta^{-1}) \right) \begin{pmatrix} e^{-i \frac43 \zeta^3-i \nu \zeta} & 0 \\ 0& e^{i\frac43 \zeta^3+i \nu \zeta} \end{pmatrix};
\]
\item[\rm (4)] $\Psi(\zeta;\nu)$ is bounded near $\zeta=0$.
\end{itemize}
\end{rhp}
This RH problem was introduced by Flaschka and Newell in \cite{FN}. They showed that from this RH problem one can retrieve the Hastings-McLeod solution for the Painlev\'e II equation. More precisely, define $q(\nu)$ by 
\begin{align*}
q(\nu)=\lim_{\zeta\to \infty} \zeta \Psi_{12}(\zeta;\nu) e^{-i \frac43 \zeta^3-i \nu \zeta}, 
\end{align*}
where $\Psi_{12}$ is the $12$-entry of $\Psi$, then $q$ is the unique solution to the Painlev\'e II equation 
\begin{align*}q''(\nu)=2q(\nu)^3+\nu q(\nu),\end{align*}
uniquely characterized by the asymptotic condition $
q(\nu)= {\rm Ai}(\nu)(1+o(1))$ as $\nu \to +\infty.$ One can show that there exists a unique solution $\Psi$ to RH problem \ref{rhp: PII rhp} if and only if the Hastings-McLeod solution $q$ has no pole at $\nu$. Since it is known that this solution has no real poles  \cite{HMcL}, it follows that $\Psi$ exists for all $\nu \in \bbR$. The kernel $K_{\rm PII}$ at the right-hand side of \eqref{eq:PII} is now given by  
\begin{equation} \label{eq: PII kernel}
K_{\rm PII}(x,y;\nu)=\frac{1}{2\pi i(x-y)} \begin{pmatrix} 1 & -1 \end{pmatrix} \Psi^{-1}(y;\nu)\Psi(x;\nu) \begin{pmatrix}1 \\ 1\end{pmatrix},
\end{equation}
where $\Psi(\zeta,\nu)$ is the unique solution to  RH problem \ref{rhp: PII rhp}. 

For interior singular point of higher order, $\rho(x)\sim (x-x^*)^{2m}$ the kernel is characterized by a RH problem associated to a special solution to the $m$-th  member of the Painlev\'e II hierarchy. 

\subsubsection{Singular edge points} 

Near a singular endpoint we have similar limits for the kernel but where  the Painlev\'e II equation is replaced by the Painlev\'e I equation. For example, in \cite{CV} it is proved that the singular endpoint with vanishing exponent $5/2$ is related to the second member of the Painlev\'e I hierarchy. For the general exponent $m+1/2$ related to  the $m$-th member of the Painlev\'e I hierarchy see \cite{CIK}.

It is interesting to note that is not  possible to obtain a singular edge point with an exponent of $3/2$ in the Hermitian one matrix model. However, by adjusting the model and considering orthogonal polynomials in the complex plane, a vanishing exponent of $3/2$ can be realized \cite{BT,DK1,FIK} and   this singular case is related to the Painlev\'e I equation.  However, there is no probabilistic interpretation for this situation.

\section{Two matrix model}
In the Hermitian two matrix model we consider the probability measure on the space of couples $(M_1,M_2)$ of $n\times n$ Hermitian matrices given  by 
\beq\label{eq:definition2m}
\frac{1}{Z_n^{2M}} \exp\left(-n  \Tr \left(V(M_1)+W(M_2)-\tau M_1M_2 \right)\right) {\rm d}M_1 {\rm d}M_2.
\eeq
Here $V$ and $W$ are two polynomials of even degree and positive leading coefficients, $Z_n^{2M}$ is a normalizing constant and $\tau >0$ is called the coupling constant. Note that if $\tau=0$, the probability measure factorizes and $M_1$ and $M_2$ are independent matrices taken randomly from \eqref{eq:onematrix} with potential $V$ and $W$ respectively.  

We will be concerned with the asymptotic behavior of the eigenvalues of $M_1$ and $M_2$ as $n\to\infty$. In this section we will discuss some general characteristics of the two matrix model and in particular the relation to certain biorthogonal polynomials.

\subsection{Master loop equation}

An important motivation for studying the two matrix model is that there is strong evidence that it generates a wide class of singular points that can not appear in the one matrix model. 

By formal calculations one can show    that the limiting eigenvalue distributions are characterized by an algebraic curve, which is also often referred to as the master loop equation and is the equivalent of \eqref{eq:masterloopequation1m} for the two matrix model. To this end, define the following functions 
\begin{align*}
Y_n(x)&=V'(x)-\frac{1}{n} \EE^{2M}_n \left[\Tr \frac{1}{x-M_1}\right]\\
X_n(y)&=W'(y)-\frac{1}{n} \EE^{2M}_n \left[\Tr \frac{1}{y-M_2}\right]\\
P_n(x,y)&=\frac{1}{n} \EE^{2M}_n \left[\Tr \frac{V'(x)-V'(M_1)}{x-M_1}\frac{W'(y)-W'(M_2)}{y-M_2}\right]\\
E_n(x,y)&=\tau (V'(x)-\tau y)(W'(y)-\tau x) -P_n(x,y)+\tau^2,
\end{align*}
where $\EE^{2M}_n$ stand for the expectation with respect to \eqref{eq:definition2m}.
The conjecture  is that these functions have expansions in $1/n^2$, i.e.
$$
Y_n=Y^{(0)}+\frac{1}{n^2} Y^{(1)}+\cdots, $$
and similarly for $X_n,P_n$ and $E_n$. Moreover, the various terms in the expansion satisfy a recursive system of equations that are  called the loop equations \cite{Eyn2} (for a survey on the formal analysis of matrix models using loop equations see for example \cite{Orantin} and the reference therein). The first of the loop equations, also called the master loop equation, reads \begin{align}\label{eq:masterloopequation2M}
E^{(0)}\left(x,\tfrac{1}{\tau}Y^{(0)}(x)\right)=E^{(0)}\left(\tfrac{1}{\tau}X^{(0)}(y),y\right)=0,
\end{align} 
which is to be interpreted as the equivalent of \eqref{eq:masterloopequation1m} for the two matrix model. 

The function $E^{(0)}$ is a polynomial of degree $\deg V$ in $x$ and degree $\deg W$ in $y$, whereas in the one matrix model \eqref{eq:masterloopequation1m} always has degree $2$ in one of the variables. Under special choices of parameters  the curve defined by \eqref{eq:masterloopequation2M} has interesting singular points that cannot occur in the one matrix model \cite{DaKaKo}.  For example, it is possible to obtain a limiting measure where the density vanishes with a vanishing exponent $p/q$. Near these points we expect to obtain new interesting scaling limits.   We recall that the only possible singular cases for the curve  \eqref{eq:masterloopequation1m} corresponding to  the one  matrix model, are the ones  listed in Section \ref{sec:sing} (and hence $q=2$).  

At this point we want to emphasize that \eqref{eq:masterloopequation2M} is derived in a formal way. In \cite{G} it was proved that the empirical eigenvalue distributions of $M_1$ and $M_2$ have weak limit almost surely. This means in particular that $Y_n$ and $X_n$ have limits, but to the best of my knowledge there is no rigorous proof of an $1/n^2$ expansion or of \eqref{eq:masterloopequation2M}.

\subsection{Biorthogonal polynomials}

The important feature of the one matrix model that made it possible to analyze it explicitly, is that it can be integrated in terms of orthogonal polynomials. In the two matrix model there is a similar structure. Let $(p_{j,n})_j$ and $(q_{k,n})_k$ be two sequences of monic polynomials, with $\deg p_{j,n}=j$ and $\deg q_{k,n}=k$, such that they satisfy the biorthogonality relation 
$$
\iint_{\bbR^2} p_{j,n}(x) q_{k,n}(y) {\rm e}^{-n\left(V(x)+W(y)-\tau x y)\right)} \ {\rm d} x {\rm d} y=h_{k,n}^2\delta_{jk}, 
$$
for certain constants $h_{k,n}^2$.  Since the orthogonality is not with respect to a Hermitian inner product, it is not a priori clear that the polynomials exists. In \cite{EMc} it was proved that they do exist, are unique and have properties that are typical for orthogonal polynomials. They have real and simple zeros \cite{EMc} and the zeros satisfy an interlacing property \cite{DGK}. Their integrable structure has been extensively explored in \cite{BEH1,BEH2,BEH}. 

As in the one matrix model \eqref{eq:1mdet}, the marginal densities or correlation function for the eigenvalues have a determinantal structure. To this end, we define \begin{align*}
P_{j,n}(y)&=\int_\bbR p_{j,n}(x){\rm e}^{-n (W(y)+V(x)-\tau x y)} {\rm d} x,\\
Q_{k,n}(x)&=\int_\bbR q_{j,n}(y){\rm e}^{-n (W(y)+V(x)-\tau x y)} {\rm d} y,
\end{align*}
and the following four kernels
\begin{align}
K_{11}^{(n)}(x_1,x_2)&=\sum_{k=0}^{n-1} \frac{1}{h_{k,n}^2} p_{k,n}(x_1) Q_{k,n}(x_2),\label{eq:2mkernel}\\
K_{12}^{(n)}(x,y)&=\sum_{k=0}^{n-1} \frac{1}{h_{k,n}^2} p_{k,n}(x) q_{k,n}(y),\nonumber \\
K_{21}^{(n)}(y,x) &=\sum_{k=0}^{n-1} \frac{1}{h_{k,n}^2} P_{k,n}(y) Q_{k,n}(x)-{\rm e}^{-n(V(x)+W(y)-\tau x y)},\nonumber \\
K_{22}^{(n)}(y_1,y_2)&=\sum_{k=0}^{n-1}  \frac{1}{h_{k,n}^2}P_{k,n}(y_1) q_{k,n}(y_2).\nonumber
\end{align}
Then by the Eynard-Mehta theorem \cite{EyM} the marginal densities of the point process given by the eigenvalues of $M_1$ and $M_2$ have the following structure
\begin{multline*}
\underbrace{\int\cdots \int}_{n-k+n-l \text{ times}} \mathcal P(x_1,\ldots,x_n,y_1,\ldots,y_n) {\rm d}x_{k+1} \cdots {\rm d}x_n {\rm d} y_{l+1} \cdots {\rm d}y_n\\
=\frac{(n-k)!(n-l)!}{n!^2} \det \begin{pmatrix}  \left(K_{11}^{(n)}(x_i,x_j)\right)_{i,j=1}^{k} & \left(K_{12}^{(n)}(x_i,y_j)\right)_{i,j=1}^{k,l}\\
\left(K_{21}^{(n)}(y_i,x_j)\right)_{i,j=1}^{l,k} &\left(K_{22}^{(n)}(y_i,y_j)\right)_{i,j=1}^{l}\end{pmatrix},
\end{multline*}
for $k,l=1,\ldots,n$.  For example, if we average over $M_2$ then we see that the eigenvalues of $M_1$ form a determinantal point process with kernel $K^{(n)}_{11}$ given in \eqref{eq:2mkernel}. To find the asymptotic behavior of the eigenavalues and the limiting local correlations, it suffices to compute the asymptotic behavior of the kernels $K^{(n)}_{ij}$.

\subsection{Riemann-Hilbert problem(s)}

The asymptotic behavior of orthogonal polynomials that appear in the one matrix model can be effectively computed using the RH problem approach (see for example \cite{Kuijlaars} for a discussion). It is therefore natural to search for a characterization of the biorthogonal polynomials in terms of a RH problem. Several such characterizations exist \cite{BEH,EMc,Kapaev,KMcL}.  Here we will discuss the RH problem from \cite{KMcL} as this was the starting point for the analysis in \cite{DG,DK2,DKM1,Mo}. It should be noted that this RH problem is equivalent to the RH problem in \cite{BEH}. 

Write $d_W=\deg W$ and define the functions
$$
w_{j,n}(x)= \int_\bbR y^j {\rm e}^{-n (W(y)+V(x)-\tau x y)} {\rm d} y, \qquad j=0,1,\ldots d_W-2.
$$ 
Then the kernel $K_{11}^{(n)}$ in \eqref{eq:2mkernel} can be characterized by the following RH problem.

\begin{rhp} \label{rhp: bio}
We look for a $d_W\times d_W$ matrix valued function $Y$ such that 
\begin{enumerate}
\item $Y$ is analytic in $\bbC \setminus \bbR$:
\item $Y_+(x)=Y_-(x)
\begin{pmatrix} 1 & w_{0,n}(x) & \cdots & w_{d_W-1,n}(x)\\
 & \ddots &&0\\ 
 & &\ddots &\\
0 & & &1
\end{pmatrix} $ for $ x\in \bbR,$
\item $Y(z)=(I+\mathcal O(1/z))\diag(z^n,z^{-n_0},\ldots,z^{-n_{d_W-2}})$ as $z\to \infty$. 
where $n_j$ is the integer part of $(n+d_W-j-2)/(d_W-1)$.  \end{enumerate}
\end{rhp}
There exists a unique solution to the RH problem \ref{rhp: bio} which can be expressed in terms of the biorthogonal polynomials \cite{KMcL}. In particular we have $Y_{11}(z)=p_{n,n}(z)$ and 
$$
K_{11}^{(n)}(x_1,x_2)= \tfrac{1}{2\pi {\rm i}(x_1-x_2)} \begin{pmatrix} 0 & w_{0,n}(x_2) & \cdots & w_{d_w-2}(x_2) \end{pmatrix} Y^{-1}_+(x_2) Y_+(x_1) \begin{pmatrix} 1\\
0\\
0\\
0\end{pmatrix} $$
For the polynomials $q_{n,n}$ and  the kernel $K_{22}^{(n)}$ a similar RH problem holds. 


The strategy for analyzing the two matrix model, is to perform a steepest descent analysis on the RH problem as $n\to \infty$.  For the general situation this is still an open problem. However, for the special case of quartic $W$ we recently analyzed the asymptotic behavior in a series of papers that we will discuss in the next section.
\section{Two matrix model with a quartic potential}

In this section we will discuss the two matrix model \eqref{eq:definition2m} for the following special choice of potentials 
\beq \label{eq:assumptionsonVW}
 \qquad V \text{ even}  \qquad \text{and} \qquad W(y)=\tfrac14y^4+\tfrac \alpha2 y^2.
\eeq
In \cite{DK2,DKM1,Mo} we  performed the steepest descent analysis for RH problem \ref{rhp: bio} for the kernel $K_{11}^{(n)}$ in \eqref{eq:2mkernel} that characterizes the eigenvalues of $M_1$. To this end, we used a vector equilibrium problem that we will discuss in Section 4.1 and 4.2.  In Section 4.3--4.5 we further assume that  $V(x)=\tfrac12x^2$ and identify all possible singular cases that can occur and present a phase diagram \cite{DGK} in Section 4.3.  In particular, there is a new type of singular point around we obtain a new kernel \cite{DG} which we discuss  in Section 4.4 and 4.5.  

\subsection{Vector equilibrium problem}

Let us assume that we are in the situation \eqref{eq:assumptionsonVW}. The key ingredient in the asymptotic analysis of \cite{
DK2,DKM1,Mo} is that we found a coulomb gas interpretation for the limiting distribution of the eigenvalues of $M_1$ that we will now describe. See also \cite{DKM2} for an alternative discussion.

For two probability measures $\mu$ and $\nu$ we define the mutual logarithmic energy $I(\mu,\nu)$ and the logarithmic energy $I(\mu)$  by \cite{SaffTotik}
$$I(\mu,\nu)=\iint \log\frac{1}{|x-y|} \ {\rm d}\mu(x){\rm d}\nu(y), \qquad I(\mu)=I(\mu,\mu).
$$
The equilibrium problem is to minimize the energy functional  $E(\nu_1,\nu_2,\nu_3)$ defined by 
\begin{multline}\label{eq:energy}
E(\nu_1,\nu_2,\nu_3)=\sum_{j=1}^3 I(\nu_j) -\sum_{j=1}^2 I(\nu_j,\nu_{j+1})\\
+\int V_1(x) \ {\rm d} \nu_1(x) + \int V_3(x) \ {\rm d} \nu_3(x), 
\end{multline}
among all vectors of measure $(\nu_1,\nu_2,\nu_3)$ satisfying the following conditions:
\begin{enumerate}\setlength{\itemsep}{0pt}
\item $\nu_1$ is a measure on $\bbR$ with total mass $1$. 
\item $\nu_2$ is a measure on ${\rm i} \bbR$ with total mass $2/3$.
\item $\nu_3$ is a measure on $\bbR$ with total mass $1/3$.
\item $\nu_2\leq \sigma_2$. \end{enumerate}
We need to clarify the external fields $V_1$ and $V_3$ acting on $\mu_1$ andÊ $\mu_3$, and the constraint $\sigma_2$ on $\mu_2$.  To this end, note that  the function 
$$s\mapsto W(s)-\tau x s$$
with $W$ as in \eqref{eq:assumptionsonVW},  has a global minimum attained at some point $s=s_1(x)$. Moreover, if $\alpha<0$ and $|x|\leq \tfrac2\tau (-\alpha/3)^{3/2}$ the function has another local minimum at some point $s_2(x)$ and a local maximum at some point $s_3(x)$.

  The external field $V_1$ in \eqref{eq:energy} is defined as 
$$
V_1(x)=V(x) +W(s_1(x))-\tau x s_1(x).
$$
Moreover, the external field $V_3$ is defined $$
V_3(x)=\begin{cases}
W(s_3(x))-\tau x s_3(x)-(W(s_2(x))-\tau x s_2(x)), & \text{if } s_{2,3}(x) \text{ exist},\\
0, & \text{otherwise}.
\end{cases}
$$
Finally,  the constraint $\sigma_2$ is a measure that is absolutely continuous with respect to the Lebesgue measure with density 
$$
\frac{{\rm d}\sigma_2(z)}{|{\rm d}z|}=\frac \tau \pi \max_{s^3+\alpha s =\tau z} \Re s, \qquad z\in {\rm i}\bbR.
$$ 
The following theorem was proved in \cite{DK2} for the case $\alpha=0$ and in \cite{DKM1}  for generalÊ $\alpha$.  See also \cite{HK}.
\begin{theorem}
Let $V$ and $W$ as in \eqref{eq:assumptionsonVW}. The energy functional $E$ defined in \eqref{eq:energy} has a unique minimizer $(\mu_1,\mu_2,\mu_2)$ among all vectors of measures satisfying the conditions  listed  below \eqref{eq:energy}. \end{theorem}
The support of the measures in the unique minimizer $(\mu_1,\mu_2,\mu_3)$ have the following structure
\[
\begin{split}
S(\mu_1)&=\cup_{j=1}^N[a_j,b_j]\\
S(\sigma- \mu_2)&= {\rm i}\bbR\setminus (-{\rm i}c_2,{\rm i}c_2),\\
S(\mu_3)&=\bbR\setminus (-c_3,c_3)
\end{split}
\]
for some $c_1,c_2>0$ and $a_1<b_1<a_2<\ldots<b_N$ and $N\in \bbN$. All measures are absolutely continuous with analytic densities (possibly  except at the origin). Moreover, if $c_2>0$ then the density of $\sigma_2-\mu_2$ vanishes as a square root at $\pm {\rm i}c_2$. Similarly for $c_3>0$ and  $\mu_3$.  

Away from the origin, the measure $\mu_1$ in the minimizer has the same behavior that one finds for the equilibrium measure in the one matrix model. Generically its density is strictly positive in the interior of the support, it vanishes as a square root at the endpoints and the variational inequality is strict. If we are in this situation and on top of that we have that in case $0\in S(\sigma-\mu_2)$ or $0\in S(\mu_3)$ then the density for that measure is strictly positive at the origin, then we say that $(V,W,\tau)$ is regular. The following result is Theorem 1.4 in \cite{DKM1}.

\begin{theorem}
Let $(V,W,\tau)$ be regular. Then, as $n\to \infty$ and $n\equiv 0  \mod 3$, the mean eigenvalue distribution of $M_1$ converges weakly to the first component $\mu_1$ of the minimizer $(\mu_1,\mu_2,\mu_3)$ of the vector equilibrium problem. 
\end{theorem}
We strongly believe that it holds also in all the singular situations and that we can drop the condition $n\equiv 0 \mod 3$ that we needed for technical reasons.

The measure $\mu_1$ describes the limiting behavior for the eigenvalues on the global scale and the natural question rises what happens on the local scale. In the regular case, the kernel $K^{(n)}_{11}$ converges to the sine kernel in the bulk and Airy kernel at the edge points as in \eqref{eq:sine} and \eqref{eq:Airy}. Away from he origin, the measure $\mu_1$ can have the same singular points as in the one matrix model and the  kernel $K^{(n)}_{11}$ has the corresponding limiting behavior. However, due to the more complicated interaction with the measure $\mu_2$ and $\mu_3$, at the origin new critical phenomena may take place which we will explore further in Sections 4.3--4.5.
\subsection{Associated Riemann surface}

At first sight, the characterization of the limiting eigenvalue distribution in terms of a vector equilibrium problem appears to be quite different from the master loop equation \eqref{eq:masterloopequation2M}. However, the solution to the vector equilibrium problem can be described  by an algebraic curve as we will now show.

Let $\mathcal R=\bigcup_{j=1}^4 \mathcal R_j$ be  where 
\[
\begin{split}
\mathcal R_1&=\overline \bbC\setminus S(\mu_1)\\
\mathcal R_2&= \bbC\setminus (S(\mu_1)\cup S(\sigma_2-\mu_2)\\
\mathcal R_3&= \bbC\setminus (S(\sigma_2-\mu_2)\cup S(\mu_3))\\
\mathcal R_4&= \bbC\setminus S(\mu_3)
\end{split}
\]
and $\mathcal R_j$ is connected to $\mathcal R_{j+1}$ in the usual crosswise manner. The following result follows from the variational conditions for the equilibrium problem and is Proposition 4.8 in \cite{DKM1}.
\begin{proposition}
The function $\xi:\mathcal R_1\to \bbC$ defined by 
\beq\label{eq:defxi}
\xi(z)=V'(z)-\int \frac{1}{z-s} \ {\rm d} \mu_1(s),
\eeq
extends to  a meromorphic function on $\mathcal R$, with a pole at infinity of degree $\mathop{deg}V-1$ on the sheet $\mathcal R_1$ and a simple pole at infinity at the other sheets.
\end{proposition}
From the last result it follows that the function $\xi$ is described by an algebraic curve
$$
\xi^4+P_3(x)\xi^3+P_2(x)\xi^2+P_1(x)\xi+ P_0(x)=0
$$
for certain polynomials $P_j(x)$. Moreover, the restriction of $\xi$ to the first sheet $\mathcal R_1$ plays the role of $Y^{(0)}$ in \eqref{eq:masterloopequation2M}.

\subsection{Phase diagram for  quartic/quadratic case}
Let us  consider the special situation \beq \label{eq:quartic/quadratic}
 \qquad V(x)=\tfrac12 x^2,  \qquad \text{and} \qquad W(y)=\tfrac14y^4+\tfrac \alpha2 y^2,
\eeq
and discuss the structure of the vector $(\mu_1,\mu_2,\mu_3)$ minimizing the energy $E$ and the associated surface $\mathcal R$. For this situation   an alternative and perhaps more direct derivation of the equilibrium  problem can be found  in \cite{DGK}.
\begin{figure}[t]
\begin{center}
\begin{tikzpicture}[scale=.9]
\draw[->](0,0)--(0,4.25) node[above]{$\tau$};
\draw[->](-4,0)--(4.25,0) node[right]{$\alpha$};
\draw[help lines] (-1,0)--(-1,1)--(0,1);
\draw[thick,rotate around={-90:(-2,0)}] (-2,0) parabola (-4.5,6.25)   node[above]{$\tau^2={\alpha+2}$};
\draw[thick]
(-1,1)..controls (0,1.5) and (-0.2,3).. (-0.1,4);
\draw[very thick]              (-1,1)..controls (-2,0.5) and (-3,0.2).. (-4,0.1) node[above]{$\alpha \tau^2 =-1$};
\filldraw  (-1,1) circle (1pt);
\draw (0.1,1) node[font=\footnotesize,right]{$1$}--(-0.1,1);
\draw (-1,0.1)--(-1,-0.1) node[font=\footnotesize,below]{$-1$};
\draw (-2,0.1)--(-2,-0.1) node[font=\footnotesize,below]{$-2$};
\draw (0.1,1.43) node [font=\footnotesize,right]{$\sqrt 2$}--(-0.1,1.43);
\draw[very thick] (2,0.8) node[fill=white]{I}
                  (-2.3,0.2) node{ IV}
                  (-2,3) node[fill=white]{III}
                  (1,3) node[fill=white]{II};
\end{tikzpicture}
\end{center}
\caption{The phase diagram in the $\alpha\tau$-plane: the critical curves $\tau^2={\alpha+2}$
and $\alpha \tau^2=-1$ separate the four cases.}
\label{fig: phase diagram}
\end{figure}

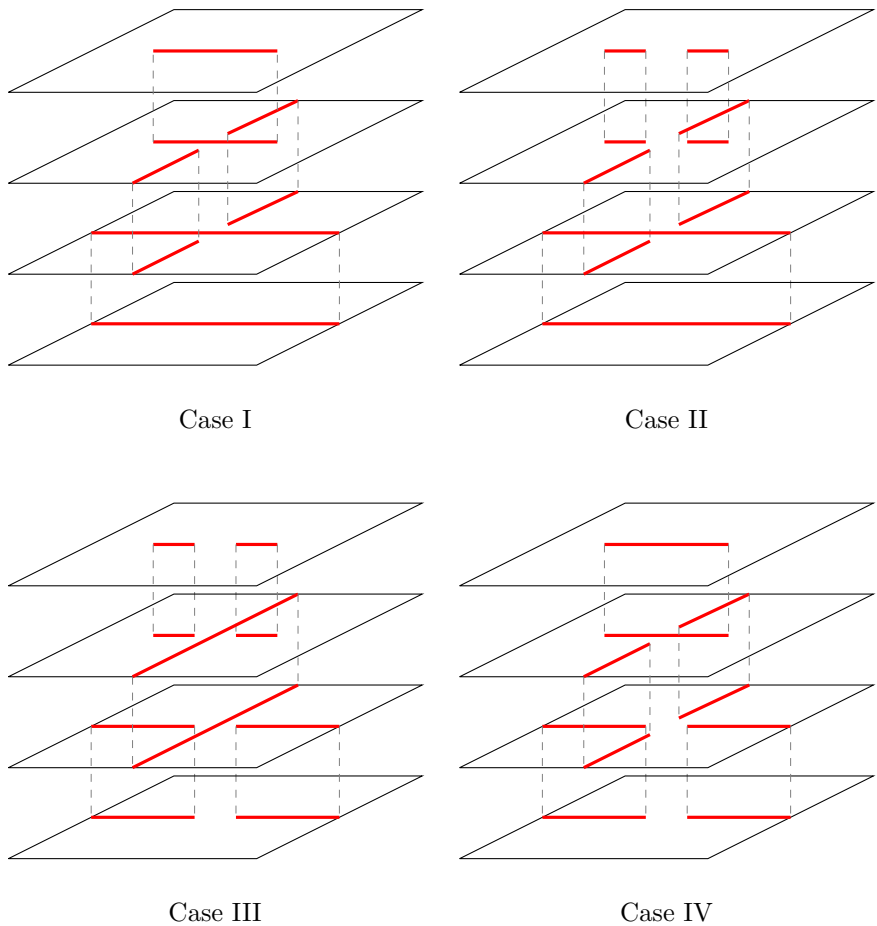
\begin{figure}
\begin{center}
\begin{tikzpicture}[xscale=1.1,yscale=1.1]
\draw (-2.75,.7) --++ (2,1)--++(3,0)--++(-2,-1)--++(-3,0);
\draw[color=red,very thick] (-1,1.2)--(0.5,1.2);
\draw (-2.75,-0.4) --++ (2,1)--++(3,0)--++(-2,-1)--++(-3,0);
\draw[color=red,very thick] (-1,.1)--(0.5,.1);
\draw[color=red,very thick] (-1.25,-.4)--(-.45,.0);
\draw[color=red,very thick] (-.1,.2)--(.75,.6);
\draw (-2.75,-1.5) --++ (2,1)--++(3,0)--++(-2,-1)--++(-3,0);
\draw[color=red,very thick] (-1.25,-1.5)--(-.45,-1.1);
\draw[color=red,very thick] (-.1,-.9)--(.75,-0.5);
\draw[color=red,very thick] (-1.75,-1)--(1.25,-1);
\draw (-2.75,-2.6) --++ (2,1)--++(3,0)--++(-2,-1)--++(-3,0);
\draw[color=red,very thick] (-1.75,-2.1)--(1.25,-2.1);
\draw[color=red,help lines,dashed] (-1,.1)--(-1,1.2);
\draw[color=red,help lines,dashed] (.5,.1)--(.5,1.2);
\draw[color=red,help lines,dashed] (-1.25,-.4)--(-1.25,-1.5);
\draw[color=red,help lines,dashed] (-.45,0)--(-.45,-1.1);
\draw[color=red,help lines,dashed] (.75,.6)--(.75,-.5);
\draw[color=red,help lines,dashed] (-.1,0.2)--(-.1,-0.9);
\draw[color=red,help lines,dashed] (-1.75,-1)--(-1.75,-2.1);
\draw[color=red,help lines,dashed] (1.25,-1)--(1.25,-2.1);
\draw (-0.25,-3.25) node {Case I};
\end{tikzpicture}\hspace{.04\textwidth}\begin{tikzpicture}[xscale=1.1,yscale=1.1]
\draw (-2.75,.7) --++ (2,1)--++(3,0)--++(-2,-1)--++(-3,0);
\draw[color=red,very thick] (-1,1.2)--(-0.5,1.2);
\draw[color=red,very thick] (0,1.2)--(0.5,1.2);
\draw[help lines,dashed] (0,1.2)--++(0,-1.1); 
\draw[help lines,dashed] (-0.5,1.2)--++(0,-1.1); 
\draw[help lines,dashed] (-1,1.2)--++(0,-1.1); 
\draw[help lines,dashed] (.5,1.2)--++(0,-1.1); 
\draw (-2.75,-0.4) --++ (2,1)--++(3,0)--++(-2,-1)--++(-3,0);
\draw[color=red,very thick] (-1,.1)--(-0.5,.1);
\draw[color=red,very thick] (0,.1)--(0.5,.1);
\draw[color=red,very thick] (-1.25,-.4)--(-.45,.0);
\draw[color=red,very thick] (-.1,.2)--(.75,.6);
\draw[help lines,dashed] (-1.25,-.4)--++(0,-1.1); 
\draw[help lines,dashed] (-0.45,0)--++(0,-1.1); 
\draw[help lines,dashed] (-0.1,.2)--++(0,-1.1); 
\draw[help lines,dashed] (.75,.6)--++(0,-1.1); 
\draw (-2.75,-1.5) --++ (2,1)--++(3,0)--++(-2,-1)--++(-3,0);
\draw[color=red,very thick] (-1.25,-1.5)--(-.45,-1.1);
\draw[color=red,very thick] (-.1,-.9)--(.75,-0.5);
\draw[color=red,very thick] (-1.75,-1)--(1.25,-1);
\draw[help lines,dashed] (1.25,-1)--++(0,-1.1); 
\draw[help lines,dashed] (-1.75,-1)--++(0,-1.1); 
\draw (-2.75,-2.6) --++ (2,1)--++(3,0)--++(-2,-1)--++(-3,0);
\draw[color=red,very thick] (-1.75,-2.1)--(1.25,-2.1);
\draw (-0.25,-3.25) node {Case II};
\end{tikzpicture}

\bigskip\bigskip

\begin{tikzpicture}[xscale=1.1,yscale=1.1]
\draw (-2.75,.7) --++ (2,1)--++(3,0)--++(-2,-1)--++(-3,0);
\draw[color=red,very thick] (-1,1.2)--(-0.5,1.2);
\draw[color=red,very thick] (0,1.2)--(0.5,1.2);
\draw[help lines,dashed] (0,1.2)--++(0,-1.1); 
\draw[help lines,dashed] (-0.5,1.2)--++(0,-1.1); 
\draw (-2.75,-0.4) --++ (2,1)--++(3,0)--++(-2,-1)--++(-3,0);
\draw[color=red,very thick] (-1,.1)--(-0.5,.1);
\draw[color=red,very thick] (0,.1)--(0.5,.1);
\draw[color=red,very thick] (-1.25,-.4)--(.75,.6);
\draw (-2.75,-1.5) --++ (2,1)--++(3,0)--++(-2,-1)--++(-3,0);
\draw[color=red,very thick] (-1.25,-1.5)--(.75,-0.5);
\draw[color=red,very thick] (-1.75,-1)--(-.5,-1);
\draw[color=red,very thick] (0,-1)--(1.25,-1);
\draw[help lines,dashed] (-.5,-1)--++(0,-1.1); 
\draw[help lines,dashed] (0,-1)--++(0,-1.1); 
\draw (-2.75,-2.6) --++ (2,1)--++(3,0)--++(-2,-1)--++(-3,0);
\draw[color=red,very thick] (-1.75,-2.1)--(-.5,-2.1);
\draw[color=red,very thick] (0,-2.1)--(1.25,-2.1);
\draw[help lines,dashed] (-1,1.2)--++(0,-1.1); 
\draw[help lines,dashed] (0.5,1.2)--++(0,-1.1); 
\draw[help lines,dashed] (-1.25,-.4)--++(0,-1.1); 
\draw[help lines,dashed] (.75,.6)--++(0,-1.1);  
\draw[help lines,dashed] (-1.75,-1)--++(0,-1.1); 
\draw[help lines,dashed] (1.25,-1)--++(0,-1.1); 
\draw (-0.25,-3.25) node {Case III};
\end{tikzpicture}\hspace{.04\textwidth}\begin{tikzpicture}[xscale=1.1,yscale=1.1]
\draw (-2.75,.7) --++ (2,1)--++(3,0)--++(-2,-1)--++(-3,0);
\draw[color=red,very thick] (-1,1.2)--(0.5,1.2);
\draw (-2.75,-0.4) --++ (2,1)--++(3,0)--++(-2,-1)--++(-3,0);
\draw[color=red,very thick] (-1,.1)--(0.5,.1);
\draw[color=red,very thick] (-1.25,-.4)--(-.45,.0);
\draw[color=red,very thick] (-.1,.2)--(.75,.6);
\draw[help lines,dashed] (-0.45,0)--++(0,-1.1); 
\draw[help lines,dashed] (-0.1,.2)--++(0,-1.1); 
\draw (-2.75,-1.5) --++ (2,1)--++(3,0)--++(-2,-1)--++(-3,0);
\draw[color=red,very thick] (-1.25,-1.5)--(-.45,-1.1);
\draw[color=red,very thick] (-.1,-.9)--(.75,-0.5);
\draw[color=red,very thick] (-1.75,-1)--(-.5,-1);
\draw[color=red,very thick] (0,-1)--(1.25,-1);
\draw[help lines,dashed] (-.5,-1)--++(0,-1.1); 
\draw[help lines,dashed] (0,-1)--++(0,-1.1); 
\draw (-2.75,-2.6) --++ (2,1)--++(3,0)--++(-2,-1)--++(-3,0);
\draw[color=red,very thick] (-1.75,-2.1)--(-0.5,-2.1);
\draw[color=red,very thick] (0,-2.1)--(1.25,-2.1);
\draw[help lines,dashed] (-1,1.2)--++(0,-1.1); 
\draw[help lines,dashed] (0.5,1.2)--++(0,-1.1); 
\draw[help lines,dashed] (-1.25,-.4)--++(0,-1.1); 
\draw[help lines,dashed] (.75,.6)--++(0,-1.1);  
\draw[help lines,dashed] (-1.75,-1)--++(0,-1.1); 
\draw[help lines,dashed] (1.25,-1)--++(0,-1.1); 
\draw (-0.25,-3.25) node {Case IV};
\end{tikzpicture}
\end{center}
\caption{The Riemann surface in the four different cases that are possible in the case $V(x)=\tfrac12x^2$ and $W(y)=\tfrac14 y^4 +\tfrac\alpha 2y^2$.} 
\label{fig:4cases}
\end{figure}

In this special situation we also have that $\mu_1$ is supported on one or two intervals. Hence there exists $a,c_1,c_2,c_3\geq 0$ such 
\[\begin{split}
S(\mu_1)&=[-a,a]\setminus(-c_1,c_1),\\
S(\sigma_2- \mu_2)&= {\rm i}\bbR\setminus (-{\rm i}c_2,{\rm i}c_2),\\
S(\mu_3)&=\bbR\setminus (-c_3,c_3).
\end{split}
\]
If $c_1=0$ the $\mu_1$ is supported on one interval and if $c_1>0$ it is supported on two intervals. We now distinguish four different cases
 \begin{center}
\begin{tabular}{llll}
{\bf Case I} & $c_1,c_3=0$ and $c_2>0$ \\{\bf Case II} &$c_3=0$ and $c_1,c_2>0$.\\
{\bf Case III}& $c_2=0$ and $c_1,c_3>0$ \\
{\bf Case IV}  &$c_1=0$ and $c_2,c_3>0$.
\end{tabular}
 \end{center}
In Figure \ref{fig:4cases} we showed the sheet structure of the associated Riemann surface $\mathcal R$ in each of the four case. Depending on the values of $\tau$ and $\alpha$ we are in one of these four cases (or a transition from one to the other). 

In  Figure \ref{fig: phase diagram} we plotted an $\alpha \tau$-phase diagram. The $\alpha\tau$-plane is separated into four regions by the curves $\tau^2={2+\alpha}$ and $\alpha \tau=-1$ and each region corresponds to a particular case.   As long as we are not on one of the two separating curves, we are in the generic situation and the local correlation for the eigenvalues of $M_1$ are given by the sine kernel at the bulk and Airy kernel at the edge.  However, if we pass one of the separating curves there is a transition  and at the origin the local correlations may have different limits. We will now discuss the various transitions but leave the multicritical situation $(\alpha,\tau)=(-1,1)$ to the next section. See also Figure \ref{fig:4casestrans} for the Riemann surface $\mathcal R$ at the points of transition.

{\bf Case I $\leftrightarrow$ Case II.} In this transition the support $S(\mu_1)$ splits from one interval into two intervals. At the splitting point, the density of $\mu_1$ vanishes quadratically. The situation is the same as in the one matrix model and again the local correlations are governed by the kernel $K_{\rm PII}$ as given in \eqref{eq: PII kernel}. 

{\bf Case IV$\leftrightarrow$ Case I.}  Here we see that there is a transition in the support of $S(\mu_3)$. In Case IV there is a gap in the support $S(\mu_3)$ that closes at the point of transition to dissapear in Case I. In fact, the transition is similar to the transition from Case I to Case II, but now it concerns the measure $\mu_3$. As it turns out, it does not have an effect on the eigenvalues of $M_1$.

{\bf Case III $\leftrightarrow$ Case IV.}  When we travel from Case III to Case IV  the two intervals in the support of $S(\mu_1)$ merge to one interval, but simultaneously a gap in the support of $S(\sigma_2-\mu_2)$ opens up. The origin is a branch point connecting the first three sheets of the Riemann surface. This situation cannot happen in the case  of the one matrix model. However, it has appeared before in the one matrix model with external source Ê\cite{BK}. The local correlation in this case are given by the Pearcey kernel
\beq\label{eq:Pearcey}
K_{{\rm Pe}}(x,y;s) =\frac{1}{(2\pi {\rm i})^2} \int_{\mathcal C} \int_{-{\rm i}\infty}^\infty {\rm e}^{\frac14 w^4-\frac s2 w^2+xw-\frac14 z^4+\frac s2 z^2-yw} 
\ \frac{{\rm d}z{\rm d}w}{z-w}, \eeq
where $\mathcal C$ is a contour that consists of two rays from $\pm \inftyÊ{\rmÊe}^{{\rm i} \pi/4}$ to $0$  together with two rays from $0$ to $\pm \inftyÊ{\rmÊe}^{-{\rm i} \pi/4}$. See also \cite{BH,BH1,OR,TW}.

{\bf Case II $\leftrightarrow$ Case III.} The situation is very similar to the transition from Case III to Case IV but now the origin is a branch point connecting the three sheets $\mathcal R_2,\mathcal R_3$ and $\mathcal R_4$. As in the transition from Case IV to Case I it does not effect the eigenvalue distribution of $M_1$.

\subsection{Critical point} \label{sec:DG}

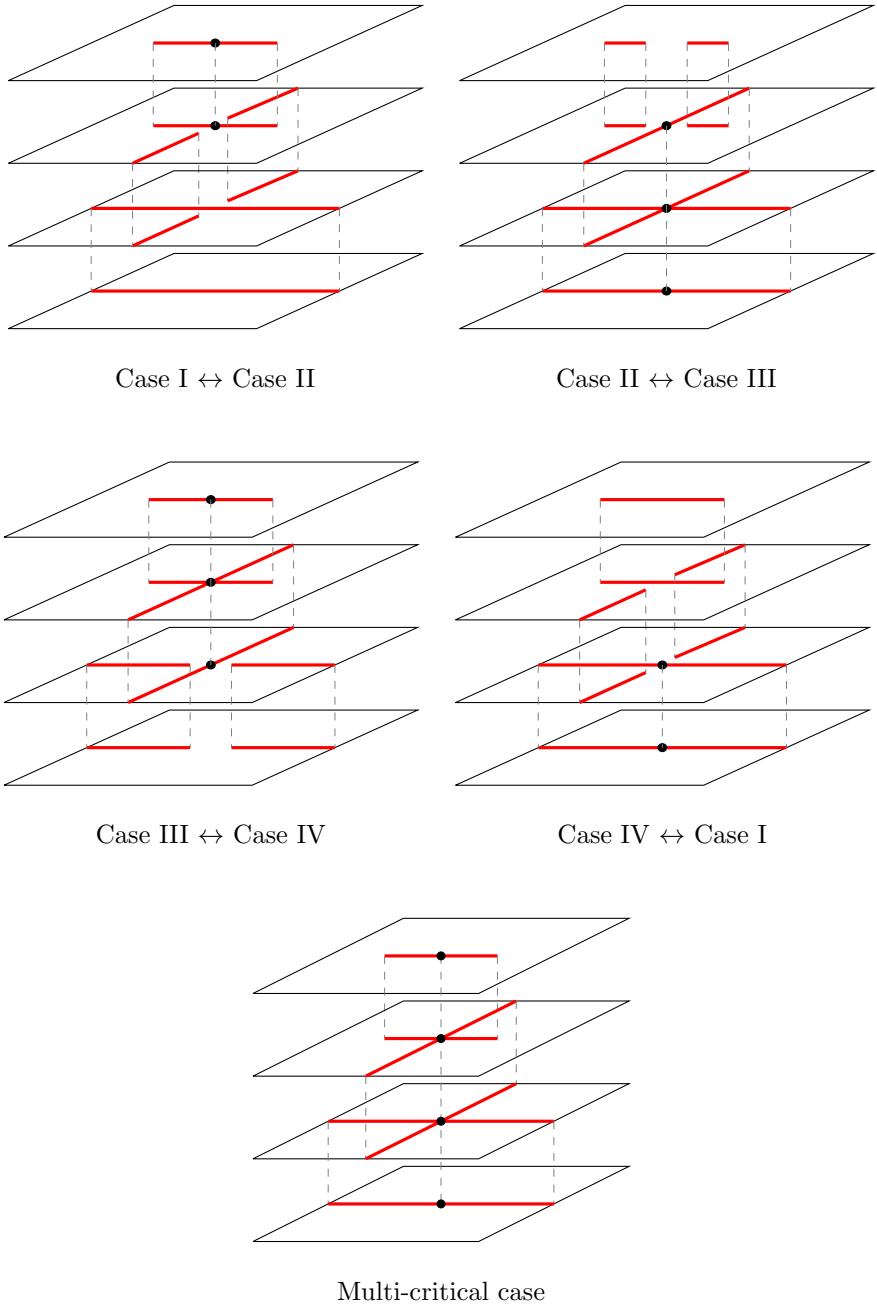
\begin{figure}
\begin{center}
\begin{tikzpicture}[xscale=1.1,yscale=1]
\draw (-2.75,.7) --++ (2,1)--++(3,0)--++(-2,-1)--++(-3,0);
\draw[color=red,very thick] (-1,1.2)--(0.5,1.2);
\draw (-2.75,-0.4) --++ (2,1)--++(3,0)--++(-2,-1)--++(-3,0);
\draw[color=red,very thick] (-1,.1)--(0.5,.1);
\draw[color=red,very thick] (-1.25,-.4)--(-.45,.0);
\draw[color=red,very thick] (-.1,.2)--(.75,.6);
\draw[help lines,dashed] (-.45,0)--++(0,-1.1); 
\draw[help lines,dashed] (-.1,.2)--++(0,-1.1); 
\draw (-2.75,-1.5) --++ (2,1)--++(3,0)--++(-2,-1)--++(-3,0);
\draw[color=red,very thick] (-1.25,-1.5)--(-.45,-1.1);
\draw[color=red,very thick] (-.1,-.9)--(.75,-0.5);
\draw[color=red,very thick] (-1.75,-1)--(1.25,-1);
\draw (-2.75,-2.6) --++ (2,1)--++(3,0)--++(-2,-1)--++(-3,0);
\draw[color=red,very thick] (-1.75,-2.1)--(1.25,-2.1);
\filldraw(-.25,1.2) circle(1.5pt);
\filldraw(-.25,.1) circle(1.5pt);
\draw[help lines,dashed] (-.25,1.2)--++(0,-1.1); 
\draw[help lines,dashed] (-1,1.2)--++(0,-1.1); 
\draw[help lines,dashed] (0.5,1.2)--++(0,-1.1); 
\draw[help lines,dashed] (-1.25,-.4)--++(0,-1.1); 
\draw[help lines,dashed] (.75,.6)--++(0,-1.1);  
\draw[help lines,dashed] (-1.75,-1)--++(0,-1.1); 
\draw[help lines,dashed] (1.25,-1)--++(0,-1.1); 
\draw (-0.25,-3.25) node {Case I $\leftrightarrow$ Case II};
\end{tikzpicture}\hspace{.04\textwidth}\begin{tikzpicture}[xscale=1.1,yscale=1]
\draw (-2.75,.7) --++ (2,1)--++(3,0)--++(-2,-1)--++(-3,0);
\draw[color=red,very thick] (-1,1.2)--(-0.5,1.2);
\draw[color=red,very thick] (0,1.2)--(0.5,1.2);
\draw[help lines,dashed] (-.5,1.2)--++(0,-1.1);
\draw[help lines,dashed] (0,1.2)--++(0,-1.1);  
\draw (-2.75,-0.4) --++ (2,1)--++(3,0)--++(-2,-1)--++(-3,0);
\draw[color=red,very thick] (-1,.1)--(-0.5,.1);
\draw[color=red,very thick] (0,.1)--(0.5,.1);
\draw[color=red,very thick] (-1.25,-.4)--(.75,.6);
\draw (-2.75,-1.5) --++ (2,1)--++(3,0)--++(-2,-1)--++(-3,0);
\draw[color=red,very thick] (-1.75,-1)--(1.25,-1);
\draw[color=red,very thick] (-1.25,-1.5)--(.75,-0.5);
\draw (-2.75,-2.6) --++ (2,1)--++(3,0)--++(-2,-1)--++(-3,0);
\draw[color=red,very thick] (-1.75,-2.1)--(1.25,-2.1);
\draw (-0.25,-3.25) node {Case II $\leftrightarrow$ Case III};
\filldraw(-.25,-1) circle(1.5pt);
\filldraw(-.25,.1) circle(1.5pt);
\filldraw(-.25,-2.1) circle(1.5pt);
\draw[help lines,dashed] (-.25,.1)--++(0,-2.2); 
\draw[help lines,dashed] (-1,1.2)--++(0,-1.1); 
\draw[help lines,dashed] (0.5,1.2)--++(0,-1.1); 
\draw[help lines,dashed] (-1.25,-.4)--++(0,-1.1); 
\draw[help lines,dashed] (.75,.6)--++(0,-1.1);  
\draw[help lines,dashed] (-1.75,-1)--++(0,-1.1); 
\draw[help lines,dashed] (1.25,-1)--++(0,-1.1); 
\end{tikzpicture}

\bigskip\bigskip

\begin{tikzpicture}[xscale=1.1,yscale=1]
\draw (-2.75,.7) --++ (2,1)--++(3,0)--++(-2,-1)--++(-3,0);
\draw[color=red,very thick] (-1,1.2)--(0.5,1.2);
\draw (-2.75,-0.4) --++ (2,1)--++(3,0)--++(-2,-1)--++(-3,0);
\draw[color=red,very thick] (-1,.1)--(0.5,.1);
\draw[color=red,very thick] (-1.25,-.4)--(.75,.6);
\draw (-2.75,-1.5) --++ (2,1)--++(3,0)--++(-2,-1)--++(-3,0);
\draw[color=red,very thick] (-1.25,-1.5)--(.75,-0.5);
\draw[color=red,very thick] (-1.75,-1)--(-.5,-1);
\draw[color=red,very thick] (0,-1)--(1.25,-1);
\draw[help lines,dashed] (-.5,-1)--++(0,-1.1); 
\draw[help lines,dashed] (0,-1)--++(0,-1.1); 
\draw (-2.75,-2.6) --++ (2,1)--++(3,0)--++(-2,-1)--++(-3,0);
\draw[color=red,very thick] (-1.75,-2.1)--(-0.5,-2.1);
\draw[color=red,very thick] (0,-2.1)--(1.25,-2.1);
\filldraw(-.25,-1) circle(1.5pt);
\filldraw(-.25,.1) circle(1.5pt);
\filldraw(-.25,1.2) circle(1.5pt); 
\draw[help lines,dashed] (-1,1.2)--++(0,-1.1); 
\draw[help lines,dashed] (0.5,1.2)--++(0,-1.1); 
\draw[help lines,dashed] (-1.25,-.4)--++(0,-1.1); 
\draw[help lines,dashed] (.75,.6)--++(0,-1.1);  
\draw[help lines,dashed] (-1.75,-1)--++(0,-1.1); 
\draw[help lines,dashed] (1.25,-1)--++(0,-1.1); 
\draw[help lines,dashed] (-.25,1.2)--++(0,-2.2); 
\draw (-0.25,-3.25) node {Case III $\leftrightarrow$ Case IV};
\end{tikzpicture}\hspace{.04\textwidth}\begin{tikzpicture}[xscale=1.1,yscale=1]
\draw (-2.75,.7) --++ (2,1)--++(3,0)--++(-2,-1)--++(-3,0);
\draw[color=red,very thick] (-1,1.2)--(0.5,1.2);
\draw (-2.75,-0.4) --++ (2,1)--++(3,0)--++(-2,-1)--++(-3,0);
\draw[color=red,very thick] (-1,.1)--(0.5,.1);
\draw[color=red,very thick] (-1.25,-.4)--(-.45,.0);
\draw[color=red,very thick] (-.1,.2)--(.75,.6);
\draw[help lines,dashed] (-.45,0)--++(0,-1.1); 
\draw[help lines,dashed] (-.1,.2)--++(0,-1.1); 
\draw (-2.75,-1.5) --++ (2,1)--++(3,0)--++(-2,-1)--++(-3,0);
\draw[color=red,very thick] (-1.25,-1.5)--(-.45,-1.1);
\draw[color=red,very thick] (-.1,-.9)--(.75,-0.5);
\draw[color=red,very thick] (-1.75,-1)--(1.25,-1);
\draw (-2.75,-2.6) --++ (2,1)--++(3,0)--++(-2,-1)--++(-3,0);
\draw[color=red,very thick] (-1.75,-2.1)--(1.25,-2.1);
\filldraw(-.25,-2.1) circle(1.5pt);
\filldraw(-.25,-1) circle(1.5pt);
\draw[help lines,dashed] (-.25,-1)--++(0,-1.1);
\draw[help lines,dashed] (-1,1.2)--++(0,-1.1); 
\draw[help lines,dashed] (0.5,1.2)--++(0,-1.1); 
\draw[help lines,dashed] (-1.25,-.4)--++(0,-1.1); 
\draw[help lines,dashed] (.75,.6)--++(0,-1.1);  
\draw[help lines,dashed] (-1.75,-1)--++(0,-1.1); 
\draw[help lines,dashed] (1.25,-1)--++(0,-1.1); 
\draw (-0.25,-3.25) node {Case IV $\leftrightarrow$ Case I};
\end{tikzpicture}
\bigskip \bigskip 

\begin{tikzpicture}[xscale=1,yscale=1]
\draw (-2.75,.7) --++ (2,1)--++(3,0)--++(-2,-1)--++(-3,0);
\draw[color=red,very thick] (-1,1.2)--(0.5,1.2);
\draw[help lines,dashed] (-1,1.2)--++(0,-1.1); 
\draw[help lines,dashed] (0.5,1.2)--++(0,-1.1); 
\draw (-2.75,-0.4) --++ (2,1)--++(3,0)--++(-2,-1)--++(-3,0);
\draw[color=red,very thick] (-1,.1)--(0.5,.1);
\draw[color=red,very thick] (-1.25,-.4)--(.75,.6);
\draw[help lines,dashed] (-1.25,-.4)--++(0,-1.1); 
\draw[help lines,dashed] (.75,.6)--++(0,-1.1);  
\draw (-2.75,-1.5) --++ (2,1)--++(3,0)--++(-2,-1)--++(-3,0);
\draw[color=red,very thick] (-1.25,-1.5)--(.75,-0.5);
\draw[color=red,very thick] (-1.75,-1)--(1.25,-1);
\draw[help lines,dashed] (-1.75,-1)--++(0,-1.1); 
\draw[help lines,dashed] (1.25,-1)--++(0,-1.1); 
\draw (-2.75,-2.6) --++ (2,1)--++(3,0)--++(-2,-1)--++(-3,0);
\draw[color=red,very thick] (-1.75,-2.1)--(1.25,-2.1);
\draw[help lines,dashed] (-.25,1.2)--++(0,-3.3); 
\filldraw(-.25,-2.1) circle(1.5pt);
\filldraw(-.25,-1) circle(1.5pt);
\filldraw(-.25,.1) circle(1.5pt);
\filldraw(-.25,1.2) circle(1.5pt);
\draw (-0.25,-3.25) node {Multi-critical case};
\end{tikzpicture}
\end{center}
\caption{The Riemann surface in the transition from one case to the other  for $V(x)=\tfrac12x^2$ and $W(y)=\tfrac14 y^4 +\tfrac\alpha 2y^2$.} 
\label{fig:4casestrans}
\end{figure}

We now deal with the multicritical point $(\alpha,\tau)=(-1,1)$. In this case, the origin connects all four sheets of the Riemann surface $\mathcal R$, see also the picture at the bottom of  Figure \ref{fig:4casestrans}.  In this case the function $\xi$ in \eqref{eq:defxi} is given by the algebraic curve
$$\xi^4-x \xi^3 +x^2=0$$
and the limiting eigenvalue distribution  vanishes with a square root near the origin which is at the \textit{interior} of the support. In \cite{DG} we characterized the limiting kernel which we will now discuss. To this end we first need the following RH problem. \begin{rhp} \label{rhp: tacnode rhp} Let $s,t\in \bbR$. We search for a $4 \times 4$ matrix-valued function $M(\zeta)$ satisfying
\begin{itemize}
\item[\rm (1)] $M$ is analytic for  $\zeta \in \bbC \setminus \Sigma_M$;
\item[\rm (2)] $M_+(\zeta)=M_-(\zeta)J_k$, for $\zeta \in \Gamma_k,$ $k=0,\ldots,9$;
\item[\rm (3)] As $\zeta \to \infty$ with $\zeta \in \bbC \setminus \Sigma_M$ we have
\begin{multline}\label{eq:asymM}
M(\zeta)=\left( I+\mathcal O(\zeta^{-1}) \right) \diag \left((-\zeta)^{-1/4},\zeta^{-1/4},(-\zeta)^{1/4},\zeta^{1/4} \right)\\   \times \small{\tfrac{1}{\sqrt 2} \begin{pmatrix} 1 & 0 & -i & 0 \\ 0 & 1& 0&i \\ -i&0&1&0\\0&i&0&1 \end{pmatrix}} 
\diag \left( e^{-\tfrac23 (-\zeta)^{3/2} -2 s (-\zeta)^{1/2}+t\zeta}, e^{-\tfrac23 \zeta^{3/2} -2 s \zeta^{1/2}-t \zeta},\right.\\\left.e^{\tfrac23 (-\zeta)^{3/2} +2 s (-\zeta)^{1/2}+t \zeta},e^{\tfrac23 \zeta^{3/2} +2 s \zeta^{1/2}-t \zeta} \right).
\end{multline}
\item[\rm (4)] $M(\zeta)$ is bounded near $\zeta=0$.
\end{itemize}

The fractional powers in   $\zeta\mapsto \zeta^{3/2}$, $\zeta\mapsto \zeta^{1/2}$ and $\zeta\mapsto \zeta^{\pm 1/4}$ are chosen such that these maps are analytic in $\bbC\setminus (-\infty,0]$ and take positive values on the positive part of the real line. The fractional powers in  $\zeta\mapsto (-\zeta)^{3/2}$, $\zeta\mapsto (-\zeta)^{1/2}$ and $\zeta\mapsto (-\zeta)^{\pm 1/4}$ are chosen such that these maps are analytic in $\bbC\setminus [0,\infty)$ and take positive values on the negative part of the real line.

The contour $\Sigma_M$ is shown in Figure \ref{fig: contour M} and consists of 10 rays emanating from the origin. The function $M(\zeta)$ makes constant jumps $J_k$ on each of the rays $\Gamma_k$. These rays are determined by two angles $\varphi_1$ and $\varphi_2$ satisfying $0<\varphi_1<\varphi_2<\pi/2$. The half-lines $\Gamma_k,$ $k=0,\ldots,9$, are defined by 
\begin{align*}
\Gamma_0 &= \bbR^+, & \Gamma_1 &= e^{i\varphi_1} \bbR^+, & \Gamma_2 &= e^{i\varphi_2} \bbR^+, \\
&&                 \Gamma_3 &= e^{i(\pi-\varphi_2)} \bbR^+, & \Gamma_5 &= e^{i(\pi-\varphi_1)} \bbR^+,    
\end{align*}
and $ \Gamma_{5+k}=-\Gamma_k,$ for $k=0,\ldots,4.$
All rays are oriented towards infinity.
\end{rhp}
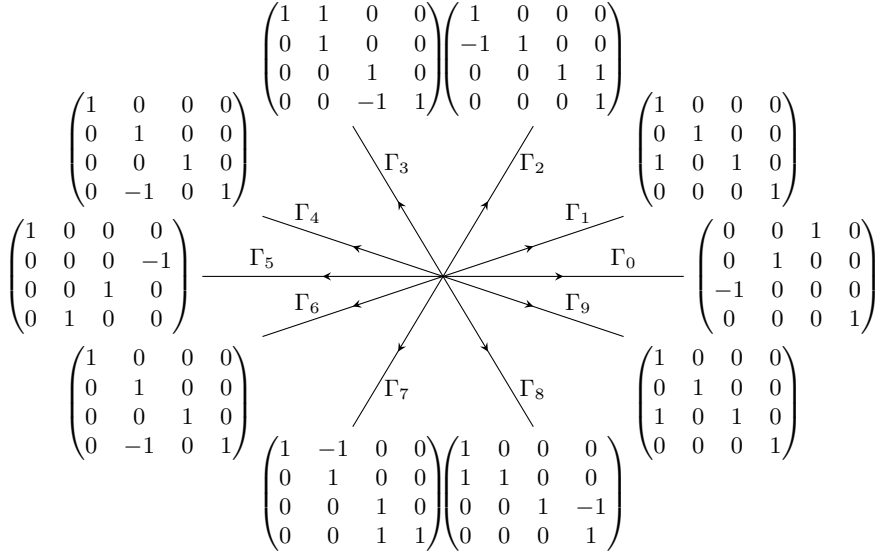
\begin{figure}[t]
\centering
\small{
\begin{tikzpicture}[scale=.8]
\begin{scope}[decoration={markings,mark= at position 0.5 with {\arrow{stealth}}}]
\draw[postaction={decorate}]      (0,0)--node[near end, above]{$\Gamma_0$}(4,0) node[right]{$\begin{pmatrix} 0&0&1&0\\0&1&0&0\\-1&0&0&0\\0&0&0&1 \end{pmatrix}$};
\draw[postaction={decorate}]      (0,0)--node[near end, above]{$\Gamma_1$}(3,1) node[above right]{$\begin{pmatrix} 1&0&0&0\\0&1&0&0\\1&0&1&0\\0&0&0&1 \end{pmatrix}$};
\draw[postaction={decorate}]      (0,0)--node[near end, right]{$\Gamma_2$}(1.5,2.5) node[above]{$\begin{pmatrix} 1&0&0&0\\-1&1&0&0\\0&0&1&1\\0&0&0&1 \end{pmatrix}$};
\draw[postaction={decorate}]      (0,0)--node[near end, right]{$\Gamma_3$}(-1.5,2.5) node[above]{$\begin{pmatrix} 1&1&0&0\\0&1&0&0\\0&0&1&0\\0&0&-1&1 \end{pmatrix}$};
\draw[postaction={decorate}]      (0,0)--node[near end, above]{$\Gamma_4$}(-3,1) node[above left]{$\begin{pmatrix} 1&0&0&0\\0&1&0&0\\0&0&1&0\\0&-1&0&1 \end{pmatrix}$} ;
\draw[postaction={decorate}]      (0,0)--node[near end, above]{$\Gamma_5$}(-4,0) node[left]{$\begin{pmatrix} 1&0&0&0\\0&0&0&-1\\0&0&1&0\\0&1&0&0 \end{pmatrix}$};
\draw[postaction={decorate}]      (0,0)--node[near end, above]{$\Gamma_6$}(-3,-1) node[below left]{$\begin{pmatrix} 1&0&0&0\\0&1&0&0\\0&0&1&0\\0&-1&0&1 \end{pmatrix}$} ;
\draw[postaction={decorate}]      (0,0)--node[near end, right]{$\Gamma_7$}(-1.5,-2.5)node[below]{$\begin{pmatrix} 1&-1&0&0\\0&1&0&0\\0&0&1&0\\0&0&1&1 \end{pmatrix}$};
\draw[postaction={decorate}]      (0,0)--node[near end, right]{$\Gamma_8$}(1.5,-2.5)node[below]{$\begin{pmatrix} 1&0&0&0\\1&1&0&0\\0&0&1&-1\\0&0&0&1 \end{pmatrix}$};
\draw[postaction={decorate}]      (0,0)--node[near end, above]{$\Gamma_9$}(3,-1)node[below right]{$\begin{pmatrix} 1&0&0&0\\0&1&0&0\\1&0&1&0\\0&0&0&1 \end{pmatrix}$};
\end{scope}
\end{tikzpicture}}
\caption{The jump contour $\Sigma_M$ in the complex $\zeta$-plane and the constant jump matrices $J_k$ on each of the rays $\Gamma_k$, $k=0, \ldots,9$.}
\label{fig: contour M}
\end{figure}

For $s,t\in \bbR$ the solution to the RH problem exist and is unique \cite[Th. 2.2]{DG}. Moreover, the RH problem is related to the Hastings-McLeod solution for the Painlev\'e  II equation. Indeed, by rewriting $I+\mathcal O(1/\zeta))=I+M^{(1)}/\zeta+\ldots$ in \eqref{eq:asymM} we have  
$$
\left(M^{(1)}\right)_{1,4}=i2^{-1/3}q\left(2^{2/3}(2s-t^2)\right),
$$
where the left-hand side is the $14$ entry of $M^{(1)}$ and  $q$ stands for the Hastings-McLeod solution. In fact, by taking derivatives with respect to $\zeta$ and the parameters $s,t$ we can obtain from the RH problem  a system of first order differential equations for $M$  for which the Painlev\'e II equation appears as the compatibility condition \cite{DKZ,DG}.

We now define $K_{\rm cr}$ by
\begin{align*}\label{eq:defKcr}
K_{\rm cr}(u,v;s,t)=
\frac{1}{2 \pi i (u-v)} \begin{pmatrix} 1 & 1 & 0 & 0 \end{pmatrix}  M({\rm i}v;s,t) ^{T}  M({\rm i}u;s,t)^{-T} \begin{pmatrix} -1\\1\\0\\0 \end{pmatrix}.
\end{align*}
where $M^T$ stands for the transpose of $M$ and $M^{-T}$ for the inverse transpose. The following  theorem is the main result in \cite{DG}. 
\begin{theorem}
Let $V$ and $W$ be as  in \eqref{eq:quartic/quadratic} and set 
\begin{equation*}
\begin{pmatrix} \alpha \\ \tau \end{pmatrix} = \begin{pmatrix} -1 \\ 1 \end{pmatrix} + a n^{-1/3}\begin{pmatrix} 2 \\ 1 \end{pmatrix} + b n^{-2/3}\begin{pmatrix} -1 \\ 2 \end{pmatrix}, 
\end{equation*}
for  $a,b\in \bbR$. Then for $n\to \infty$ and $n\equiv 0 \mod 6$, and $K_{11}^{(n)}$ as in \eqref{eq:2mkernel} we have \begin{equation} \label{eq:DGkernel}
\lim_{n \to \infty} \frac{1}{n^{2/3}}K_{11}^{(n)}\left(\frac{u}{n^{2/3}},\frac{v}{n^{2/3}}\right) =K_{\rm cr}\left(u,v;\tfrac14 (a^2-5b),-a\right),
\end{equation}
uniformly for $u,v$ in compact subsets of $\bbR$.
\end{theorem}

 It is interesting that RH problem \ref{rhp: tacnode rhp} with $t=0$ appeared before \cite{DKZ} in the characterization of the kernel near a tacnode singularity in a model of non-intersecting brownian paths (see also   \cite{AFM,Jtac}) for alternative characterizations). The apparent reason for this is that also in the case of the tacnode singularity, the critical measure vanishes as a square root near an interior point. Nevertheless,  the kernel describing the tacnode singularity is constructed in a different way out of the RH problem. In \cite{DG} we proved that, perhaps somewhat surprisingly,  the two kernels define essentially different processes.

\subsection{Reductions of the new kernel}

There is a a certain hierarchy in the limiting kernels that  we have discussed so far. Let us denote the sine and Airy  kernel (i.e. the right-hand sides of \eqref{eq:sine} and \eqref{eq:Airy}) by $K_{\rm sine}$ and $K_{\rm Ai}$.  We also recall the definition of $K_{\rm PII}$ and $K_{\rm Pe}$ in \eqref{eq:  PII kernel} and \eqref{eq:Pearcey}.  The kernel $K_{\rm cr}$ is on top of the hierarchy, in the sense that the four other kernels are limit points for $K_{\rm cr}$.  
For instance, from the phase diagram  in Figure \ref{fig: phase diagram} we see that if we walk from the critical point $(\alpha,\tau)=(-1,1)$ to the right along the curve $\tau^2=2+\alpha$, we  end up in the in the critical situation described by the kernel $K_{\rm PII}$. In \cite{DG} we  proved that $K_{\rm PII} $ is a limit point of $K_{\rm cr}$ in the following way.
There exists a function $h$ such that 
\[
\lim_{ a \to +\infty} 2^{\frac53} a \frac{h(x,a)}{h(y,a)}K_{\rm cr} \left(2^{\frac53} ax,2^{\frac53} ay;\frac{a^2}{2},-a\left(1-\frac{\sigma}{a^2} \right)\right)=K_{\rm PII}(x,y;2^{\frac53}\sigma),
\]
uniformly for $x,y$ in compact sets. (Note that as the correlation functions are determinantal they are invariant under conjugation of the kernel by the function $h$ and the point process is not changed)

If we walk from $(\alpha,\tau)=(-1,1)$ along the curve $\tau=-1/\alpha$ in the phase diagram, then we expect to obtain the Pearcey kernel as a limit. In a recent paper \cite{GZ} the authors proved that 
\begin{align}
\lim_{a\to +\infty} \frac{1}{\sqrt 2 a^\frac14} K_{\rm cr} \left(\frac{x}{\sqrt 2 a^\frac14},\frac{y}{\sqrt 2 a^\frac14},-\tfrac12 a^2,a\left(1-\frac{\sigma}{2 a^\frac32}\right)\right)=K_{\rm Pe}(x,y;\sigma),
\end{align}
uniformly for $x,y$ in compact subsets. 

Finally, we note that the transitions $K_{\rm Pe}\to K_{\rm Ai}$ and $K_{\rm Ai}\to K_{\rm Sine}$ can  for example be proved by classical steepest descent methods on the integral representations for the Airy and Pearcey functions. Moreover, the transition $K_{\rm PII}\to K_{\rm Ai}$ can be found using steepest descent techniques on the RH problem \ref{rhp: PII rhp} for $K_{\rm PII}$.

\section{Singular points in one matrix models revisited}

Now that we have characterized the correlations for the eigenvalues of $M_1$ it is natural to ask about the eigenvalues of $M_2$, especially for the multi-critical point in the quartic/quadratic case. We will now answer this question in a general fashion and show that the multi-critical point discussed in Section 4 can be embedded into a  larger class  for which a rigorous treatment may be within reach.

First note in case  $$\text{$V(x)=\tfrac{1}{2}x^2$ and $W$ arbitrary,}$$ we can complete the square  and rewrite \eqref{eq:definition2m} as 
\beq
\frac{1}{Z_n^{2M}} \exp\left(-n  \Tr \left( \tfrac12(M_1-\tau M_2)^2+W(M_2)-\tfrac12 \tau^2M_2 \right)\right) {\rm d}M_1 {\rm d}M_2.
\eeq
Hence the matrices $M_1-\tau M_2$ and  $M_2$ are independent matrices taken randomly  from the  GUE  (=Gaussian Unitary Ensemble) and  the one matrix model \eqref{eq:onematrix} with potential $$W_{\rm eff}(y)=W(y)-\tfrac12 \tau^2y^2.$$  Moreover, by writing the matrix $M_1$ as  $M_1=(M_1-\tau M_2)+\tau M_2,$ we see that $M_1$   is a linear combination of two random matrices taken independently from the GUE and the one matrix model \eqref{eq:onematrix} with potential $W_{\rm eff}$ respectively. 

It is also important to note that in this situation  the spectral curve \eqref{eq:masterloopequation2M} can also be computed in a fairly explicit way, 
\beq \label{eq:spectralcurveinterpol}
\tau (x-\tau \xi ) (W_{\rm eff} '(\xi)+\tau^2 \xi -\tau x) - Q(\xi)=0,
\eeq 
where $Q$ is the polynomial
$$
Q(\xi)=\int \frac{W_{\rm eff}'(\xi)- W_{\rm eff}'(t)}{\xi-t} {\rm d}\mu_{W_{\rm eff}}(t),
$$ 
and $\mu_{W_{\rm eff}}$ is the equilibrium problem measure minimizing \eqref{eq:energyexternalfield} with potential $W_{\rm eff}$. We recall that there is no proof that \eqref{eq:spectralcurveinterpol} indeed characterizes the limiting eigenvalue distribution of the matrix $M_1$,  but we are free to use it as an ansatz in a  RH analysis and prove this fact a posteriori. This may be a good starting point of dealing with some particular interesting examples, some of which we will now discuss.

 Let us  proceed by choosing the parameters in the potential $W$ so that $W_{\rm eff}$ does not depend on $\tau$ and consider the  family of random matrices 
$$M(\tau)= M_{\rm GUE} +\tau M_{W_{\rm eff}}$$ parametrized by $\tau$. Note that  $(\frac{1}{1+\tau}M(\tau))_\tau$ interpolates between $M_{\rm GUE}$Ê for  $\tau=0$ andÊ $M_{W_{\rm eff}}$ for  $\tau\to \infty$.  Let us  also assume that  $W_{\rm eff}$ leads to an equilibrium measure $\mu_{W_{\rm eff}}$ for which the density vanishes with exponent $2m$ at the origin. Then on the local scale we expect to see a transition  from the sine kernel  to the  kernel associated to the singular interior point (related to the Painlev\'e II hierarchy).

To see how this transition comes about we rewrite \eqref{eq:spectralcurveinterpol} to \beq \nonumber \label{eq:spectralcurveinterpol2}
\left(\tau (x-\tau \xi )-\frac12 (W_{\rm eff} '(\xi)\right)^2 =\tfrac{1}{4} W_{\rm eff}'(\xi)^2- Q(\xi)=-\xi^{4m}(c+\mathcal O(\xi)),\eeq
as $\xi \to 0$, for some positive constant Ê$c>0$. Set $\tau_{\rm cr}=\sqrt{-W_{\rm eff} ''(0)/2}$. If $\tau\neq \tau_{\rm cr}$ then for $x=0$ we find that $\xi=0$ is a double solution and we have a  singular  point that connects two sheets   of the Riemann surface. If the point is on the physical sheet $\mathcal R_1$ then the local correlations are  given by the same  kernel as we have for $M_{W_{\rm eff}}$. In case $\tau=\tau_{cr}$ then for $x=0$ the solution $\xi=0$ is of multiplicity at least four (in case $m=1$) or at least six (in case $m\geq 2)$ and hence the singular point connects several sheets. Here we expect a more complicated kernel, characterized by a RH problem that has the same size as the number of sheets involved. 

Concluding, by adding  an independent GUE matrix to a matrix taken randomly from a one matrix model with a singular interior point, we can construct higher order critical phenomena.  It is an  interesting open problem to analyze these cases explicitly and identify the local correlations (and their reductions).   \\[-6pt]

\noindent\textit{Example 1.} In the first example we return to the  case in Section \ref{sec:DG} and consider
\beq
V(x)=\tfrac12 x^2, \qquad \text{and} \qquad W(y)=\tfrac14y^4-\tfrac{2-\tau^2}{2}y^2,\eeq
so we are are on the curve $\alpha=-2+\tau^2$ in  the phase diagram in Figure \ref{fig: phase diagram}.  Hence we have $W_{{\rm eff}}(y)=y^4/4-y^2$  and hence, in a double scaling limit, the local correlation for the eigenvalues for $M_2$  are governed by the Painlev\'e II kernel \eqref{eq: PII kernel}. As for the eigenvalues of $M_1$, from the  phase diagram we read off that for $\tau<1$ the local correlation are given by the sine kernel and for $\tau>1$ we obtain the Painlev\'e II kernel. The transition takes place at $\tau=1$ (and hence $\alpha=-1$) where the kernel has the scaling limit \eqref{eq:DGkernel}. \\[-6pt]

\noindent \textit{Example 2.} In the second example we consider the $$
W(y)= \tfrac{1}{6}y^6 - \tfrac{1}{2^{4/3}} y^4  +\tfrac{1}{2} (-\tfrac{1}{2^{1/3}} + \tau^2) y^2  
$$
For this potential the equilibrium measure for  $W_{\rm eff}$ is given by
$$
\frac{{\rm d} \mu_{W_{\rm eff}}}{{\rm d}x}=\tfrac{4}{\pi} x^4\sqrt{2^{2/3}-x^2},$$
and we have a quartic vanishing at the origin. In this case, the spectral curve \eqref{eq:spectralcurveinterpol2} takes the form
$$
2^{-4/3}\xi^2 + \xi^4+ 
 \tau (x - \tau \xi) \left(\tau x + 2^{1/3} \xi - \tau^2 \xi + 2^{2/3} \xi^3 - \xi^5\right)=0,
 $$ In particular, for $\tau=2^{-1/3}$ we have
$$  x^2 + \xi^6 - x \xi^3 ( 2^{1/3} \xi^2-2)=0,
$$ 
  and from here we see that for $x=0$ the solution $\xi=0$ has multiplicity six and the origin connects all sheets.\\[-6pt]
  
  \noindent \textit{Example 3.} In the final example we show that similar phenomena occur when dealing with singular edge points. Consider the potential 
$$
W(y)= \tfrac{8}{5} y + (\tfrac{1}{5} + \tfrac{\tau^2}{2}) y^2 - \tfrac{4}{15} y^3  +\tfrac{1}{20} y^4.
$$
In this case, the effective potential $W_{\rm eff}$ gives rise to an equilibrium measure with a singular endpoint
$$
\frac{{\rm d} \mu_{W_{\rm eff}}}{{\rm d}x}=\tfrac{1}{10\pi}(x-2)^2\sqrt{4-x^2}.$$
See also \cite{CV}.  For the special case $\tau=\sqrt 5/5$ the spectral curve takes the form 
$$
(-32 + 16 y - 8 y^2 + 4 y^3 - 
   y^4) + \sqrt 5(8 + 4  y - 4 y^2 + y^3) x - 
 5 x^2=0
$$
For $x=4 \sqrt 5 /5$ we find  four solutions $\xi=2$ and hence this point connects all sheets. 

\end{document}